# Possible atomic structures for the sub-bandgap absorption of chalcogen hyperdoped silicon

Ke-Fan Wang,[1,2] Hezhu Shao,[3] Kong Liu,[2] Shengchun Qu,[2,a)] Yuanxu Wang,[1,a)] Zhanguo Wang[2]

[1]*Institute for Computational Materials Science, School of Physics and Electronics, Henan University, Kaifeng 475004, P. R. China*

[2]*Key Laboratory of Semiconductor Materials Science, Institute of Semiconductors, Chinese Academy of Sciences, Beijing 100083, P. R. China*

[3]*Ningbo Institute of Material Technology and Engineering, Chinese Academy of Sciences, Ningbo 315201, P.R. China*

Single-crystal silicons were hyperdoped respectively by sulfur, selenium, and tellurium element using ion implantation and nanosecond laser melting. The hyperdoping of such chalcogen elements endowed the treated silicon with a strong and wide sub-bandgap light absorption. When these hyperdoped silicons were thermally annealed even at low temperatures (such as 200~400 ℃), however, this extra sub-bandgap absorption began to attenuate. In order to explain this attenuation process, we consider it corresponding to a chemical decomposition reaction from optically absorbing structure to non-absorbing structure, and obtain a very good fitting to the attenuated absorption coefficient by using Arrhenius equation. Further, we

[a)] Authors to whom correspondence should be addressed equally. Electronic mail: qsc@semi.ac.cn (Prof. S. Qu) or wangyx@henu.edu.cn (Prof. Y. Wang);

extract the reaction activation energies from the fittings and they are 0.343±0.031 eV for S-, 0.426±0.042 eV for Se-, and 0.317±0.033 eV for Te-hyperdoped silicon, respectively. We discuss these activation energies in term of the bond energies of chalcogen-Si metastable bonds, and finally suggest that several high-energy interstitial sites instead of the substitutional site, are very possibly the atomic structures that are responsible for the strong sub-bandgap absorption of chalcogen hyperdoped silicon.

In the last decade years, single-crystal silicon hyperdoped by chalcogen elements (S, Se or Te) using ultrafast laser techniques has drawn more and more research attentions, due to its novel optoelectronic properties especially the strong and wide sub-bandgap (>1100 nm) light absorption[1,2]. This additional infrared absorption favors the feasibility to fabricate silicon-based high-efficiency solar cells and room temperature infrared photodetectors. In fact, such optoelectronic devices have been produced but still with a not satisfactory performance at the sub-bandgap wavelength[3-6]. In order to improve the sub-bandgap photoelectric response of these devices, firstly we should know the atomic structure that can bring the strong sub-bandgap absorption, and then can we control and utilize it. *Unfortunately, this interesting atomic structure and its deactivating process upon thermal annealing are both still controversial at present.*

Previous researchers[7-10] have found from first-principles calculations that single doped chalcogen atom in substitutional site has the minimum energy and it also can introduce a deep and filled impurity band into the energy gap, which make it naturally to be considered as the candidate for the sub-bandgap absorption. However, this supposition didn't acquire a reliable confirmation from the fitting result of extended X-ray absorption fine structure (EXAFS) spectroscopy for non-annealed Se hyperdoped silicon[11]. Further, if it was the substitutional site that brings the sub-bandgap absorption mostly, then this extra absorption would be little affected by the thermal annealing because that the substitutional site is already the stablest structure, which situation would contradict the experimental findings[2]. *So the substitutional site may not be the atomic structure that is responsible for the sub-bandgap absorption of chalcogen hyperdoped silicon.*

On the other hand, Simmons et al [12] considered that the optical deactivation of S-hyperdoped silicon during thermal annealing is due to a diffusion-mediated change of chemical state, and obtained a diffusion activation energy of 1.7±0.1 eV that could explain their experimental results very well. Finally they suggested that the deactivation was caused by a transformation for sulfur impurities from an isolated substitutional site (optically absorbing) to dimmers or complexes (optically non-absorbing). In their scenario, the isolated S atom has to diffuse about 1.5 nm or longer to find and combine with other S atoms. At low anneal temperatures such as 200~400℃, the sub-bandgap absorption of chalcogen hyperdoped silicon already begins to decrease[2, 12-14], but the corresponding diffusion lengths for S, Se and Te atoms inside single-crystal silicon are nominally about 0.1 nm, 0.01 nm and 0.01 nm[13], respectively, one or two order lower than the required 1.5 nm or longer. Also, the obvious Se diffusion was not observed by Secondary Ion Mass Spectrometry (SIMS) for the Se-hyperdoped silicon annealed at 1100 K for 100 s[12]. *As a result, the drop of sub-bandgap absorption for chalcogen hyperdoped silicon after thermally annealed may not be attributed to the thermal diffusion of chalcogen atoms.*

One of the authors, Shao et al[10] have investigated sixteen typical metastable chalcogen-S atomic structures by first-principles calculations, and found that among them, eight quasi-substitutional and four interstitial structures can introduce impurity bands into the silicon forbidden band, and after thermal annealing, these high-energy chalcogen-S metastable structures transform to be three low-energy two-coordinated interstitial structures that cannot produce any defect level in the band gap. Additionally, the calculated coordination number (2) and bond length (2.23~2.31 Å) of these three low-energy interstitial structures agree very well with the EXAFS fitted results of Se-hyperdoped silicon annealed at 1100 K or 1225 K for 30 min[11]. *Consequently, we believe that the deactivation of sub-bandgap absorption may be due to the local chemical conversion for the high-energy chalcogen-Si atomic structures.*

In this letter, firstly we report that S-, Se-, and Te-hyperdoped silicon samples were prepared by ion implantation plus pulsed laser melting, and then they were thermally annealed at different temperatures (200~800 ℃) for 30 minutes. The sub-bandgap absorptances of the annealed samples were measured and the corresponding absorption coefficient can be well fitted by considering the conversion from optically absorbing state to non-absorbing state as a chemical decomposition reaction. This fitting can be used to extract the reaction activation energies that are discussed in term of the bond energy of chalcogen-Si bond.

B-doped p-type, 200-μm-thick Si wafers (1~10 Ω·cm, double-side polished, Czochralski-grown, (100) orientation) were cleaned successively by isopropanol, acetone and ethanol in order to remove the surface organic contaminants. Then the wafers were ion implanted at room temperature with 95 keV $^{32}S^+$, or 176 keV $^{80}Se+$, or 245 keV $^{130}Te^+$ at 7° off-normal to prevent channeling effect in a dose of $1\times10^{16}$ cm$^{-2}$. Subsequently, the ion implanted wafers were laser melted by a spatially homogenized pulsed KrF excimer laser beam (248 nm, 20 ns duration full width at half maximum, rectangle in size 4 mm×1 mm). The used laser fluences for melting S-, Se- and Te-implanted silicon wafers were 1.7, 2.3 and 2.3 J/cm$^2$, respectively. Each point on the sample surface was irradiated by an average of four pulses. Finally, the hyperdoped silicon wafers were cleaved into 1cm × 1cm square pieces for thermal annealing that was carried out in a horizontal tube furnace at different temperatures for 30 minutes, flushed continuously with a high purity $N_2$ gas.

The total hemispherical (specular and diffuse) reflectance (R) and transmittance (T) of the samples were measured using an AvaSpec-2048 UV-VIS-NIR spectrometer over the wavelength range from 1100 nm to 2500 nm. The spectrometer was equipped with an integrating sphere detector. The total absorptance (A = 1- T- R) of the samples was determined from the directly measured T and R. The instruments were calibrated by sets of Labsphere Diffuse Reflectance Standards.

The first-principles calculations of chalcogen hyperdoped silicon are performed by the DFT[15, 16] method within the generalized gradient approximation with the Perdew, Burke, and Ernzerhof (PBE) function[17], as implemented in the Vienna ab initio simulation package (VASP)[18, 19], which employs a plane-wave basis. In our theoretical work, crystal model is adopted, whose framework is a 3×3×3-$Si_8$ cubic supercell containing one isolated chalcogen atom to obtain a dilution level similar with that reported in experiments. Owing to the highly non-thermal equilibrium conditions in preparation of the chalcogen-hyperdoped silicon, we construct various geometric configurations by putting chalcogen atoms into different positions in silicon and then relax them. More details about the first-principles calculation can be found elsewhere[10].

We measured the sub-bandgap reflectances and transmittances of the annealed hyperdoped silicons, and then obtained their sub-bandgap absorptances, as shown in Fig 1. Thermal annealing can reduce all of the sub-bandgap absorptances of S, Se or Te hyperdoped silicons. But the dependence of this reducing on the anneal temperature is a little different between the respective case of S, Se or Te hyperdoping. For S hyperdoped silicon, its absorptance begins to drop at the anneal temperature of 200 °C and disappears completely after thermally annealed at 700 °C for 30 min. For Se hyperdoped silicon, its absorptance starts to reduce when it was annealed at 400 °C, and attenuates almost completely at 800 °C. For Te hyperdoped silicon, its absorptance commences decreasing at 400 °C and after annealed at 800 °C, the absorptance still remains about 15%.

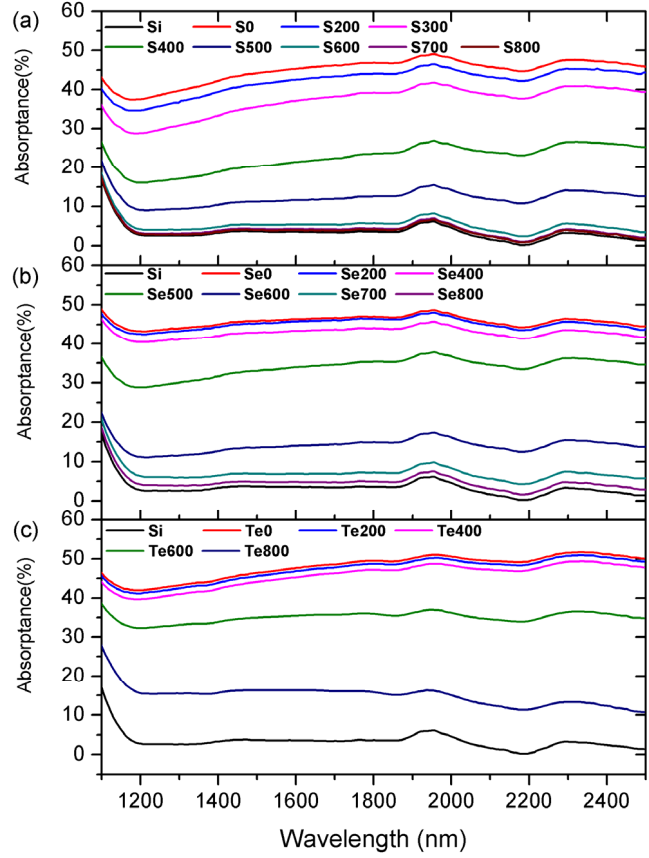

FIG. 1. Sub-bandgap absorptances of S- (a), Se- (b) and Te- (c) hyperdoped silicon samples annealed for 30 minutes at the temperatures indicated in the legend.

In order to analyze these attenuated absorptances further, we plot their values at 1800 nm as a function of *kT*, as shown in Fig 2(a), where *k* is the Boltzmann constant and *T* is the absolute anneal temperature (in unit of Kelvin). Inspired by the exponent-like shape of the decreased absorptance and the above analysis on deactivation mechanisms, we suppose that the transformation from optically absorbing state (denoted by "A" below) to non-absorbing state (denoted by "B" below) corresponds to a chemical decomposition reaction, and then write the reaction equation as follows,

$$A \xrightarrow{K} B \qquad (1)$$

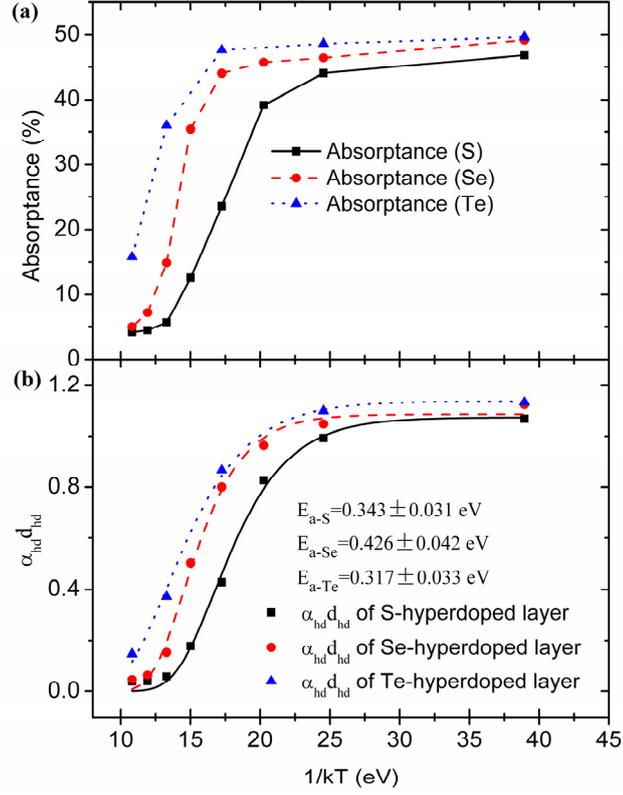

FIG. 2. The absorptance (a) and the product $\alpha_{hd} \cdot d_{hd}$ (b) of annealed samples at 1800 nm as a function of $1/kT$. Here $\alpha$, $d$, $k$ and $T$ represent the absorption coefficient and the thickness of hyperdoped layer, the Boltzmann constant and the anneal temperature (in Kelvin), respectively. Lines in (b) denote the fitting results of the attenuated $\alpha_{hd} \cdot d_{hd}$ by using Arrhenius equation. Extracted activation energies $E_a$ from the fittings are also shown in the legend of (b).

The concentration of state "A" at the moment $t$ is marked by $C_t$, which should obey such a relation,

$$-\frac{dC_t}{dt} = KC_t \tag{2}$$

where $K$ is the reaction rate constant. To solve this differential equation, we obtain,

$$K = \frac{1}{t}\ln\frac{C_0}{C_t} \tag{3}$$

where $C_0$ is the concentration of state "A" at the initial moment. The absorption coefficient should be proportional to the concentration of optically absorbing state "A" inside the hyperdoped layer. Then equation (3) converts to be,

$$K = \frac{1}{t}\ln\frac{\alpha_0}{\alpha_t} \tag{4}$$

where $\alpha_t$ and $\alpha_0$ are the sample's absorption coefficient at the moment $t$ and at the initial moment, respectively.

In view of the dependence of infrared absorptance on the anneal temperature shown in Fig 1, we believe that the chemical decomposition reaction (1) is a thermal activation process whose reaction rate constant should obey the Arrhenius equation, then

$$K = A\exp(-E_a/kT) \tag{5}$$

where $A$ is the pre-exponential factor, $E_a$ is the thermal activation energy, $k$ is the Boltzmann constant and $T$ is the absolute reaction temperature.

To combine equation (4) and (5), we obtain,

$$\alpha_t = \alpha_0 \exp(-At\exp(-E_a/kT)) \tag{6}$$

In order to facilitate the fitting below, we multiply both sides of equation (6) by the thickness of hyperdoped layer $d_{hd}$. Then the equation (6) converts to be,

$$\alpha_t d_{hd} = \alpha_0 d_{hd} \exp(-At\exp(-E_a/kT)) \tag{7}$$

To obtain the product $\alpha_{hd}d_{hd}$ of the chalcogen hyperdoped layer, we adopt the method used by J. Olea et al[20] when they dealt with the Ti-hyperdoped silicon layer, where they took into account the interface reflection between the hyperdoped layer and air at the front face, as well as that between the Si substrate and air at the back face, but neglected the interface reflection between the hyperdoped layer and the Si substrate, same as Pan et al did [21]. Then, the measured transmittance and reflectance of the hyperdoped silicon samples can be given by [20],

$$T = \frac{(1-r_{hd})(1-r_{Si})\exp(-\alpha_{hd}d_{hd})\exp(-\alpha_{Si}d_{Si})}{1-r_{hd}r_{Si}\exp(-2\alpha_{hd}d_{hd})\exp(-2\alpha_{Si}d_{Si})} \quad (8)$$

$$R = r_{hd} + \frac{(1-r_{hd})^2 r_{Si}\exp(-\alpha_{hd}d_{hd})\exp(-2\alpha_{Si}d_{Si})}{1-r_{hd}r_{Si}\exp(-2\alpha_{hd}d_{hd})\exp(-2\alpha_{Si}d_{Si})} \quad (9)$$

where

$r_{hd}$ is the reflectance of interface between air and the hyperdoped layer,

$r_{Si}$ is the reflectance of interface between air and the Si substrate,

$\alpha_{hd}$ is the absorption coefficient of the hyperdoped layer,

$\alpha_{Si}$ is the absorption coefficient of the Si substrate,

$d_{hd}$ is the thickness of the hyperdoped layer, and

$d_{Si}$ is the thickness of the Si substrate.

According to the refractive index $n$ (3.4578) and extinction coefficient $k$ (~0) of crystalline silicon at 1800 nm [22], we obtain the values of $r_{Si}$ (0.5513) and $\alpha_{Si}$ (0) using the well-generalized formula[23]. To use the values of $r_{Si}$ and $\alpha_{Si}$, and to substitute the $T$ and $R$ by the measured transmittance and reflectance, we numerically solve the equation (8) and (9) and obtain the values of $\alpha_{hd}d_{hd}$ of the different annealed hyperdoped layers and plot them as a function of $1/kT$, as shown in Fig. 2(b). Then we use the equation (7) to fit these $\alpha_{hd}d_{hd}$ points in Fig. 2(b) and obtain a very good fitting curve, which indicates that our suppositions above are reasonable. Moreover, we extracted the Ea of these chemical reactions from the fittings and they are 0.343±0.031 eV for S-, 0.426±0.042 eV for Se-, and 0.317±0.033 eV for Te-hyperdoped silicon, respectively.

The activation energy $E_a$ of a chemical reaction denotes the height of energy barrier between the reactant (here, the optically absorbing state) and the transition state, which subsequently transforms to be the product (here, the optically non-absorbing state). First-principles molecular-dynamics (MD) calculations[10] have showed that during thermal annealing of chalcogen-doped silicon, chalcogen atom breaks the previous chemical bonds before it forms new more stable bonds. Hence, we consider that the activation energy obtained above corresponds to the bond energy of chalcogen-Si chemical bonds in high-energy optically absorbing structures.

We have theoretically investigated the atomic geometry and static energy of sixteen metastable chalcogen-Si atomic structures[10]. Basing on the obtained formation energies, we try to obtain the chalcogen-Si bond energies. For a test, we calculate the Si-Si bond energy in crystalline silicon following the method as follows: the formation energy (-994.7709 eV) of the completely relaxed 3×3×3-$Si_8$ cubic supercell is divided by the amount of Si-Si bond (432), and the obtained Si-Si bond energy is -2.3027eV, whose absolute value agrees very well with that (2.3098 eV) calculated from the experimental formation enthalpy of 445.668kJ/mol at 0 K.[24] Then we use the same method to calculate the chalcogen-Si bond energy at 0K in the substitutional and interstitial atomic structures that can introduce the sub-bandgap absorption, as listed in the fourth column of TABLE 1. Fig 3 shows the corresponding X-Si (X=S, Se or Te) metastable atomic structures shown in the second column of TABLE 1. For simplicity, the Si-Si bond energy in the relaxed chalcogen-Si supercell uses the value obtained from the relaxed pure Si supercell. In addition, the chalcogen-Si bond energies in *quasi-substitutional* sites are temporally not calculated because that their configurations are very complex and usually have a higher formation energy.

TABLE I. Chalcogen-Si bond energies in substitutional and interstitial metastable atomic structures inside chalcogen hyperdoped silicon.

| Doped elements and Ea(eV) | Atomic structure | Chalcogen-Si bond length (Å), and number | Chalcogen-Si bond energy at 0 K (eV) |
|---|---|---|---|
| S 0.343±0.031 | $S_s^1$ | 2.48, 4 | -1.9523 |
| | $S_s^3$ | | |
| | $S_i^4$ | 2.12, 1 / 2.25, 2 | -2.1057 |
| | $S_i^5$ | 2.36, 4 / 2.57, 2 | -0.9760 |
| | **$S_i^6$** | **2.18, 2** | **-0.4604** |
| | $S_i^7$ | 2.41, 4 | -0.0978 |
| Se 0.426±0.042 | $Se_s^1$ | 2.57, 4 | -1.5064 |
| | $Se_i^3$ | | |
| | $Se_i^4$ | 2.25, 1 / 2.37, 2 | -1.2721 |
| | **$Se_i^5$** | **2.45, 4 / 2.64, 2** | **-0.5368** |
| | $Se_i^6$ | 2.33, 2 | 0.7012 |
| | $Se_i^7$ | 2.51, 4 | 0.5778 |
| Te 0.317±0.033 | $Te_s^1$ | 2.68, 4 | -1.2877 |
| | $Te_i^3$ | 2.48, 4 | -0.0063 |
| | $Te_i^4$ | 2.43, 1 / 2.54, 2 | -0.7533 |
| | **$Te_i^5$** | **2.60, 4 / 2.75, 2** | **-0.2011** |
| | $Te_i^6$ | 2.54, 2 | 1.3953 |
| | $Te_i^7$ | 2.66, 4 | 1.1352 |

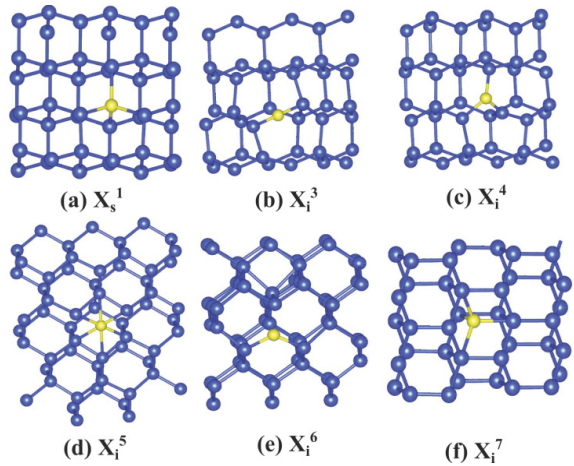

(a) $X_s^1$  (b) $X_i^3$  (c) $X_i^4$
(d) $X_i^5$  (e) $X_i^6$  (f) $X_i^7$

FIG. 3. Six metastable atomic structures with a chalcogen (X=S, Se, or Te) atom situated at a substitutional (a) or interstitial (b-f) positions in silicon. The yellow and blue spheres represent the chalcogen and silicon atoms, respectively.

From the Si-S bond energies of silicon sulfide (SiS)[24], we find that its value at 0 K decrease only about 11% for that at 800 K. Accordingly, we expect that the Si-S bond energy in sulfur hyperdoped silicon at anneal temperatures approximates that at 0 K. Since the temperature dependence of Se-Si (or Te-Si) bond energy is still unavailable, we suppose that it is similar with that of Si-S bond energy in view that S, Se and Te belong to the same group IV in periodic table and thus the similar bonding properties with Si.

For the S-Si bond energy shown in TABLE 1, we find that its absolute value in $S_i^6$ structure (-0.4604 eV) is close to the obtained activation energy (0.343±0.031 eV) for S-hyperdoped silicon, which suggests that the $S_i^6$ state is very probably the atomic structure that introduces the strong sub-bandgap absorption. The little energy difference (~0.12 eV) between them may originate from the approximations during the calculation process. The S-Si bond energy in $S_s^1$ structure is -1.9523 eV whose absolute value is much larger than the obtained activation energy (0.343±0.031 eV), which indicates that the substitutional site is not the atomic structure we are interested in, consistent with our analysis above. The S-Si bond energies in $S_i^4$, $S_i^5$ and $S_i^7$ states are all different largely from the obtained activation energy and so these atomic structures are less possible to be the origin for sub-bandgap absorption. For the same reason, the atomic structure $Se_i^5$ and $Te_i^5$ are very probably the configurations for the sub-bandgap absorption in Se-hyperdoped and Te-hyperdoped silicon, respectively.

It is worth noting that the three suggested chalcogen-Si atomic structures ($S_i^6$, $Se_i^5$ and $Te_i^5$) are only possible candidates responsible for the strong sub-bandgap absorption, due to the reasons as follows. Firstly, they are chosen from the previously calculated substitutional and interstitial structures, and so the investigation is limited. Secondly, the chalcogen-Si bond energies are roughly calculated basing on some approximations, and thus they are less accurate. Thirdly, the obtained bond energies at the structures such as $X_i^4$, $X_i^5$, are that average different length of chalcogen-Si bonds, and so it may not be the actual bond energy of the dissociated chalcogen-Si bond. So we cannot exclude the atomic structures $S_i^4$, $S_i^5$, $Se_i^4$, $Te_i^4$ being the origin for sub-bandgap absorption. More accurate identification of the optically absorbing chalcogen-Si structure needs a complete search of metastable atomic structure with sub-bandgap absorption, and then a nudged elastic band (NEB) calculation[25] should be used for each metastable structure to obtain the transformation activation energy, which are collated with the experimental values obtained here.

In summary, chalcogen-hyperdoped sing-crystal silicon samples were prepared using ion implantation and nanosecond laser melting. Then these hyperdoped samples were thermally annealed at different temperatures and their sub-bandgap absorptances were measured. By considering the attenuation of sub-bandgap absorption as a chemical decomposition reaction, we obtain the reaction activation energy which is considered to correspond to the bond energy of dissociated chalcogen-Si bond. According to the rough calculation, we suggest that three interstitial atomic structures are very possible the main origin for the strong and wide sub-bandgap absorption. However, more accurate certification of the optically absorbing chalcogen-Si atomic structure inside chalcogen hyperdoped silicon needs a complete theoretical search for all of the metastable structures and a NEB calculation for each structure to obtain the transformation activation energy.

This work was supported by the National Natural Science Foundation of China (No. 61204002, 11404348, 11305046 and U1404619), and partly by the National Basic Research Program of China (No. 2014CB643503).